\documentclass{pas}
\usepackage[normalem]{ulem}
\usepackage{amsmath}
\let\gtrsim\undefined
\let\lesssim\undefined
\usepackage{amssymb,microtype,siunitx,booktabs}
\usepackage{graphicx}	
\usepackage{hyperref}
\usepackage{cleveref}
\usepackage{multirow}
\usepackage{xspace}
\usepackage{float, makecell, array}
\usepackage{soul}
\usepackage{lmodern}
\usepackage{pdflscape}
\usepackage{rotating}
\usepackage{comment}
\usepackage[table]{xcolor}

\newcommand{\rr}{} 

\newcommand{\des}{DES\xspace}

\newcommand{\ozdes}{OzDES\xspace}

\newcommand{\hbeta}{H$\beta$\xspace}
\newcommand{\mgii}{Mg\textsc{ii}\xspace}
\newcommand{\civ}{C\textsc{iv}\xspace}

\begin{document}

\lefttitle{Publications of the Astronomical Society of Australia}
\righttitle{Cambridge Author}

\jnlPage{1}{4}
\jnlDoiYr{2025}
\doival{10.1017/pasa.2026.10226, DES-2025-891, FERMILAB-PUB-25-0823-PPD}

\articletitt{Research Paper}

\title{\ozdes Reverberation Mapping Program: C\textsc{\textmd{iv}} lags from six years of data}

\author{
\gn{Andrew} \sn{Penton}$^{1}$, \gn{Hugh} \sn{McDougall}$^{1}$, \gn{Tamara M.} \sn{Davis}$^{1}$, \gn{Zhefu} \sn{Yu}$^{2}$, \gn{Umang} \sn{Malik}$^{3}$, \gn{Paul} \sn{Martini}$^{4, 5}$, \gn{Brad E.} \sn{Tucker}$^{3}$, \gn{Christopher} \sn{Lidman}$^{6, 3}$, \gn{Geraint F.} \sn{Lewis}$^{7}$, \gn{Rob} \sn{Sharp}$^{3}$, \gn{Michel} \sn{Aguena}$^{8, 9}$, \gn{Sahar} \sn{Allam}$^{10}$, \gn{Felipe} \sn{Andrade-Oliveira}$^{11}$, \gn{Jacobo} \sn{Asorey}$^{12}$, \gn{David} \sn{Bacon}$^{13}$, \gn{Sebastian} \sn{Bocquet}$^{14}$, \gn{David} \sn{Brooks}$^{15}$, \gn{Ryan} \sn{Camilleri}$^{1}$, \gn{Aurelio} \sn{Carnero Rosell}$^{16, 9, 17}$, \gn{Daniela} \sn{Carollo}$^{8}$, \gn{Anthony} \sn{Carr}$^{18}$, \gn{Jorge} \sn{Carretero}$^{19}$, \gn{Ting-Yun} \sn{Cheng}$^{20}$, \gn{Luiz} \sn{da Costa}$^{9}$, \gn{Maria Elidaiana} \sn{da Silva Pereira}$^{21}$, \gn{Juan} \sn{De Vicente}$^{22}$, \gn{Shantanu} \sn{Desai}$^{23}$, \gn{Spencer} \sn{Everett}$^{24}$, \gn{Juan} \sn{Garcia-Bellido}$^{25}$, \gn{Karl} \sn{Glazebrook}$^{26}$, \gn{Daniel} \sn{Gruen}$^{14}$, \gn{Gaston} \sn{Gutierrez}$^{10}$, \gn{Samuel} \sn{Hinton}$^{1}$, \gn{Devon L.} \sn{Hollowood}$^{27}$, \gn{Klaus} \sn{Honscheid}$^{4, 28}$, \gn{Kyler} \sn{Kuehn}$^{29}$, \gn{Ofer} \sn{Lahav}$^{15}$, \gn{Sujeong} \sn{Lee}$^{30}$, \gn{Marisa} \sn{March}$^{31}$, \gn{Jennifer} \sn{Marshall}$^{32}$, \gn{Juan} \sn{Mena-Fern\'andez}$^{33}$, \gn{Ramon} \sn{Miquel}$^{34, 19}$, \gn{Justin} \sn{Myles}$^{35}$, \gn{Robert} \sn{Nichol}$^{36}$, \gn{Ricardo} \sn{Ogando}$^{37}$, \gn{Andrés A.} \sn{Plazas Malagón }$^{2, 38}$, \gn{Anna} \sn{Porredon}$^{22, 39}$, \gn{Martin} \sn{Rodriguez Monroy}$^{25, 40}$, \gn{Kathy} \sn{Romer}$^{41}$, \gn{Eusebio} \sn{Sanchez}$^{22}$, \gn{David} \sn{Sanchez Cid}$^{22, 11}$, \gn{Mathew} \sn{Smith}$^{42}$, \gn{Eric} \sn{Suchyta}$^{43}$, \gn{Molly} \sn{Swanson}$^{44}$, \gn{Vinu} \sn{Vikram}$^{45}$, \gn{Noah} \sn{Weaverdyck}$^{46, 47}$
}

\affil{Affiliations are listed after the references.
}

\corresp{Tamara~Davis, Email: tamarad@physics.uq.edu.au}

\citeauth{Penton et al., OzDES Reverberation Mapping Program: CIV lags from six years of data. {\it Publications of the Astronomical Society of Australia} {\bf 00}, 1--12. https://doi.org/10.1017/pasa.2026.10226}

\history{(Received 8 April 2026 UTC; revised 4 June 2026 UTC; accepted 11 June 2026 UTC)}
\begin{abstract}
We present 29 successfully recovered \civ time lags in Active Galactic Nuclei from the complete Dark Energy Survey Reverberation Mapping campaign.  The AGN in this sample span a redshift range of $1.9<z<3.5$. 
We successfully measure the velocity dispersion from the \civ spectral linewidth for 25 of these 29 sources, and use these to calculate new high-redshift black hole mass estimates, finding masses between 0.8 and 1.3 billion solar masses.  We also identify a selection effect due to the duration of the survey that can impact the radius-luminosity relation derived from    this and other (high-redshift) data.  This paper represents the culmination of the OzDES \civ campaign. 
\end{abstract}


\maketitle

\section{Introduction}
\label{intro}

Supermassive black holes (SMBHs) are thought to exist in {\rr in the centres of} all large galaxies. These extreme objects, some of which appear to be more massive than ten billion solar masses \citep{Wu2015,Onken2020,Eilers2023,Lai23,MeloCarneiro2025}, 
act as tracers of cosmic history and as drivers of galaxy evolution. Understanding how the mass of these SMBHs grows over time offers insight into feedback during galaxy formation \citep{Kormendy_2013}, the properties of the first generation of stars, galaxy merger rates \citep{Sahu2019}, and the potential existence of primordial black holes \citep{Volonteri_2021}. 

Locally, the masses of SMBHs can be inferred from the kinematics of the material orbiting the black hole but, beyond the local universe, limits on the angular resolution of our telescopes precludes the direct imaging of these dynamics.\footnote{Although Very Large Telescope Interferometer observations may enable us to probe the kinematics of the broad line region and measure the masses  of especially massive black holes \citep{Wolf24}.} Instead, the dominant method of measuring SMBH mass beyond $z=0.1$ is reverberation mapping \citep[RM;][]{Blandford1982}, a technique focusing on active galactic nuclei (AGN), which are SMBHs with active accretion disks. 

In RM, long-term monitoring is carried out simultaneously for photometric and spectroscopic observations of a {\rr quasar}, i.e.\ a {\rr highly luminous} AGN that exhibits broad emission lines. The accretion disk dominates the photometric observations, while the distinctive widening of the broad line region (BLR) emission lines allows its light to be identified via spectroscopy, and so we can track the variable activity of both regions independently. Variations in the accretion disk flux are `echoed' in the variations of the broad line region, but with a lag associated with the time taken for the light to travel from the disk to the BLR, with a longer lag indicating a physically larger system. This `reverberation lag,' identified by comparing the two light curves, allows us to use time-domain measurements as a spatial ruler.
 
RM is a time-domain technique that requires costly long-term observation campaigns, but it acts as the foundation for the single-epoch method that requires only one measurement of an AGN's luminosity and spectrum {\rr \citep{Vestergaard2006}}. The single epoch method is centred around the Radius-Luminosity ($R-L$) relationship, usually expressed as a power law,  between the radius of the broad line region and the overall luminosity of the AGN. 

The $R-L$ relation is currently best constrained for low-redshift hydrogen emission lines \citep[e.g. \hbeta and H$\alpha$; ][]{HBETA_Bentz_2014}. 
Much of the history of RM has focused on low-redshift AGN, using hydrogen emission lines for a small, focused sample of high signal-to-noise sources \citep[e.g.][]{Bentz2009a}. Such sources are limited in sample count and redshift, as sources beyond $z\approx0.65$ are redshifted such that the \hbeta line falls outside the optical band. High energy reverberation emission lines have been investigated as alternatives for distant sources, including \mgii at intermediate redshifts \citep[e.g.][]{MgII_Zajacek_2020} and \civ for high-redshift sources \citep[e.g.][]{kaspi2021s, Lira2018}. RM with the \civ line has previously been investigated in low-redshift sources where the line appears in ultraviolet (UV) \citep{CIV_Peterson_2005, Metzroth2006, 2015ApJ...806..128D}. They found that the \civ lags are smaller than the lags measured from \mgii or \hbeta by a factor of about two \citep{Lira2018, Kaspi2007}.

In the last decade, RM surveys have expanded to an `industrial scale', in which hundreds to thousands of sources are observed out to high redshift. The Australian Dark Energy Survey \citep[OzDES;][]{Yuan2015, Childress2017,OZDES-DR2-Lidman_2020} contained one such industrial scale RM project, obtaining reverberation mapping measurements for {\rr $771$} 
AGN in the redshift range $z\in [0.1,3.8 ] $ over a 6-year period, with spectroscopy spanning the optical wavelength range of $\lambda \in [3750 \AA ,8800 \AA]$. \ozdes acts as a contemporary counterpart to the Sloan Digital Sky Survey \citep[SDSS;][]{shen2015}, who released their final RM sample and analysis of over $1000$ AGN sources in \citet{SDSS-Shen_2024}.

This paper acts as an extension to the OzDES work with the \civ  sample by \citet{Hoormann2019}, and a companion to the OzDES \mgii and \hbeta line analyses of \citet{yu_2023} and \citet{malik2023}. An upcoming paper will present the complete OzDES reverberation mapping sample, compare the results from the different lines, and merge with other data from the literature \citep{McDougall2025_OzDES}. In this work we present the \civ lags resulting from the full six years of monitoring. Our sources are drawn from the high-redshift ($1.6<z<4.5$) \ozdes data. They are fit with the lag recovery program \texttt{JAVELIN} \citep{Zu2011}, which assumes the light curve follows a damped random walk and uses Markov chain Monte Carlo (MCMC) to recover the lag using a uniform lag prior \citep[for details see ][]{McDougall2025_OzDES}. We compare this result to an alternative method, the Interpolated Cross-Correlation Function \citep[ICCF; ][]{ICCF-Gaskell_1987} as implemented in \texttt{PyCCF} \citep{PyCCF}. ICCF does a simple linear interpolation between data points and cross-correlates the photometric and spectroscopic light curves. Consistency between these methods is one of the criteria we use to identify high-quality lag recoveries. The resulting fits are subjected to stringent quality controls to remove contaminating false positives. In total, we find $29$ \civ AGN lags, with these successful recoveries spanning the redshift range $1.9<z<3.5$. 

The paper is structured as follows.  In \S\ref{sec:data_and_sample} we describe the observing program.  In \S\ref{sec:rec_stats} we provide an overview of the properties of the $305$ \civ sources that \ozdes monitored and outline the selection criteria we use to reduce false positives. In \S \ref{sec:BHM} we use the reverberation lags to constrain the central black hole masses for $25$ of the sources (those for which we also successfully measured the \civ linewidth). In \S\ref{sec:biases} we present an important selection bias that excludes long lags because of the limited duration of the survey, which can impact the derived radius-luminosity relation. In \S\ref{sec:conclusion} we conclude and discuss the implications of these new mass estimates on the broader topic of SMBH demographics at high redshift.  

\section{Data and Sample Selection}
\label{sec:data_and_sample}

Over the course of five years, DES imaged ten fields  with approximately weekly cadence in the $g, r, i, z$ bands, which \ozdes followed up with spectroscopy at monthly cadence for six years \citep{OZDES-DR2-Lidman_2020}.   \des imaging was performed on the CTIO Blanco 4-metre Telescope in Chile, while \ozdes spectroscopy used the Anglo-Australian Telescope's AAOmega spectrograph fed by the 2dF fibre positioner \citep[for details see][]{OZDES-DR2-Lidman_2020}.  While \des and \ozdes were designed to discover supernovae and measure redshifts of their host galaxies, the program also had very good cadence for gathering AGN continuum light curves and spectroscopic response light curves needed for RM measurements.  However, the seasonal observational window was not ideal for RM, with observations lasting roughly six months of the year for \des and five months for \ozdes; this leaves large seasonal gaps in the RM light curves.

The 2dF instrument enables 400 spectra to be taken simutaneously, and about a quarter of its fibres were placed on AGN in each exposure.   \ozdes monitored a total of $771$ AGN over six years (2013-2019) {\rr with redshifts between $0.127 < z < 3.451$}, gathering between $18$ and $25$ epochs on each AGN with a spectral resolution of $1400<R<1700$ and a wavelength range of $3700\AA$\ to $8800 \AA$.  The images and catalogues from the whole DES survey are available at \href{https://des.ncsa.illinois.edu/releases/dr2}{https://des.ncsa.illinois.edu/releases/dr2}. The OzDES spectra are available from \href{https://docs.datacentral.org.au/ozdes/overview/dr2/}{https://docs.datacentral.org.au/ozdes/overview/dr2/}. 

\begin{figure*}
    \centering
    \vspace*{-10mm}
    \makebox[\linewidth]{\includegraphics[trim={0 1cm 0 3cm},clip,scale=0.85]{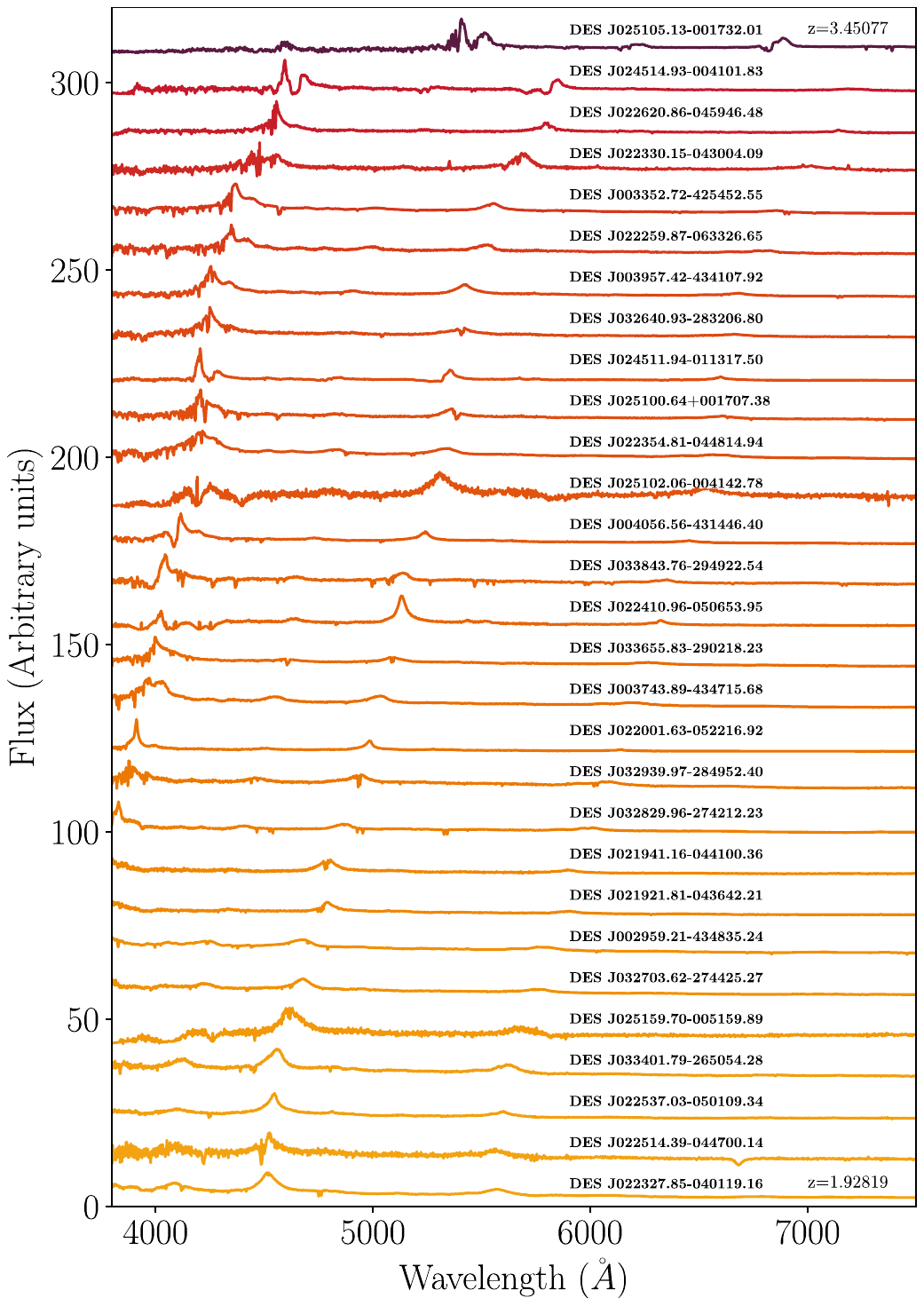}}
    \caption[{\rr Mean} spectra for all successfully recovered \civ sources.]{{\rr Mean} spectra for all successfully recovered \civ sources. Note that the \ozdes spectrum spans the wavelength range 3800-8800\r{A}, however, for visibility the plotting range has been restricted to 3800-7500\r{A}. The \civ line appears at $\sim$4537\r{A} in the lowest redshift source 
    and progresses toward the right as redshift increases (darker red indicates higher redshift). Some spectra show absorption features in the \civ line, which were severe enough in four sources to prevent an accurate linewidth measurement (see Table~\ref{tab:BHM_results}). }
    \label{fig:all_good_spectra}
\end{figure*}

We use the calibration procedures outlined in \citet{Hoormann2019} for spectroscopic measurements. We take \des photometry in the $g$, $r$ and $i$ bands (discarding any of low quality, lacking adequate calibration, or outliers), and then average measurements that were taken on the same night. These are used to calibrate the \ozdes spectra for each source, correcting for wavelength-dependent losses due the finite size of the fibres, and these calibrated spectra are then used to identify reverberating emission lines, their variations in strength for RM, and their velocity dispersions for SMBH mass estimates; see \citet{Hoormann2019} for more detail. {\rr The OZDES spectra of all 29 AGN with measured CIV lags are shown in \Cref{fig:all_good_spectra}.}

\begin{figure*}
    \centering
    \includegraphics[scale=0.4]{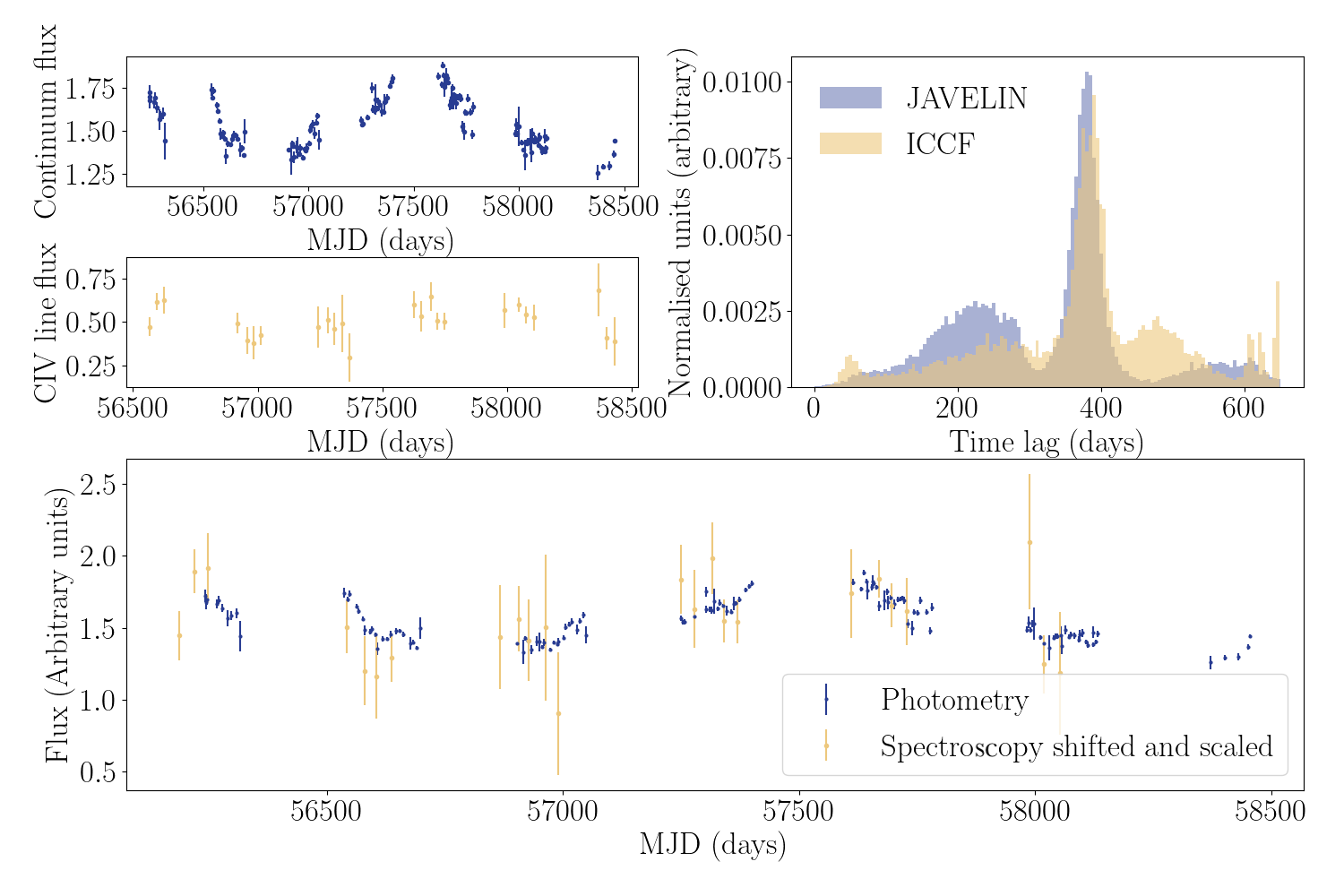}
    \caption[Gold sample example.]{An example of gold rated source, DES~J022620.86-045946.48.  In the photometric and spectroscopic lightcurves we can see that there is a long term smooth variation that is present in both lightcurves. This is an important feature in most high quality lag measurements. This leads to a posterior for the lag (top right) with a sharp peak in both \texttt{JAVELIN} and ICCF. The smaller broader peaks at $\sim$180 days and $\sim$540 days correspond to seasonal gaps and are likely caused by aliasing. }
    \label{fig:goldexample}
\end{figure*}

\begin{figure*}
    \centering
   \includegraphics[scale=0.4]{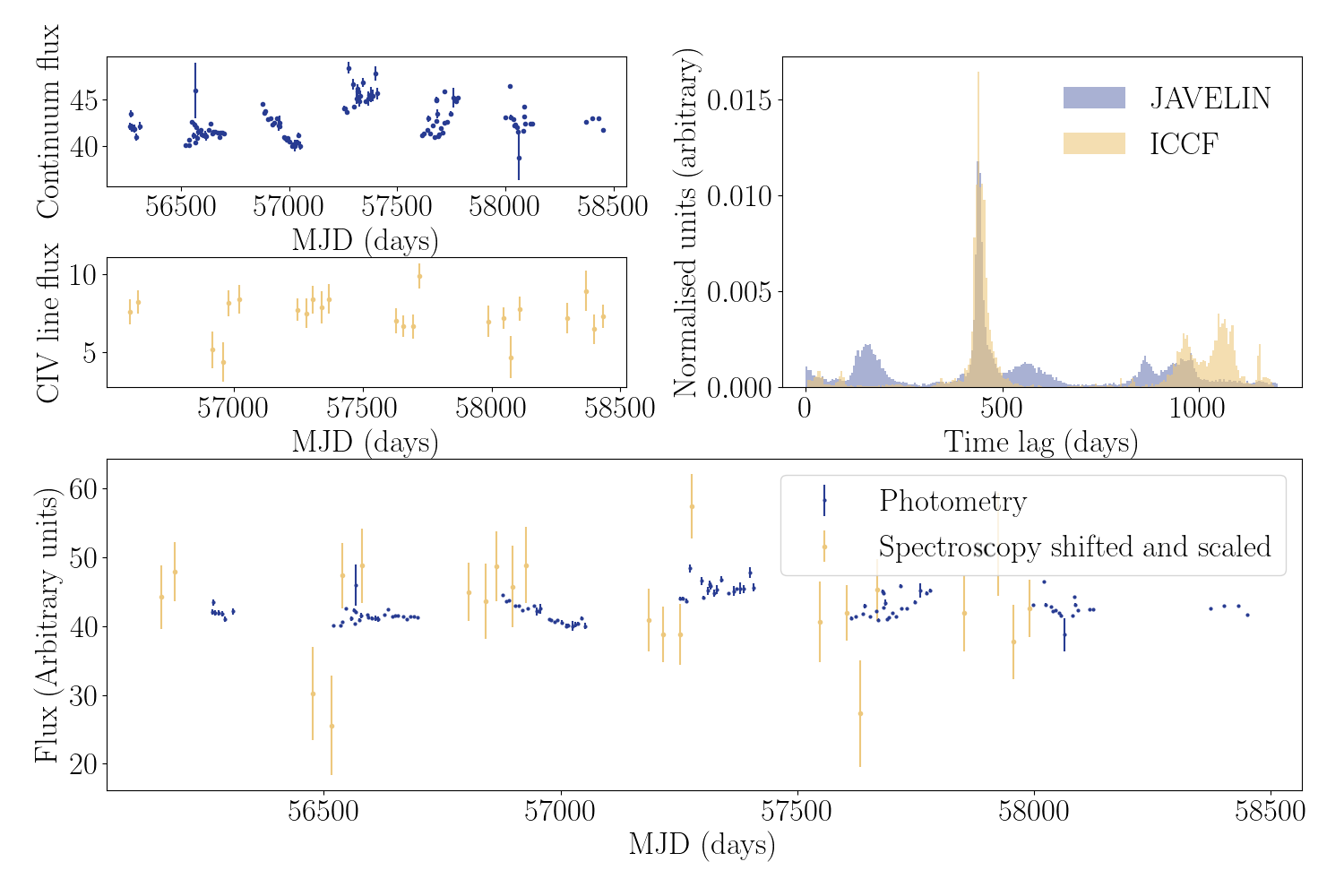}
    \caption[Silver sample example.]{An example of silver rated source, DES~J002959.21-434835.24. In the original photometric and spectroscopic lightcurves we can see that there is a long term variation present in the photometric lightcurve, however in this case there is a less obvious signal in the spectroscopic lightcurve. This leads to a posterior for the lag with a sharp peak in both \texttt{JAVELIN} and ICCF accompanied by many smaller peaks. These smaller peaks are likely caused by aliasing given their location, however, since there is much more of the probability contained within them, this example is rated as having a lower quality lag measurement than the example shown in  \Cref{fig:goldexample}. }
    \label{fig:silverexample}
\end{figure*}

\begin{figure*}
    \centering
     \includegraphics[scale=0.4]{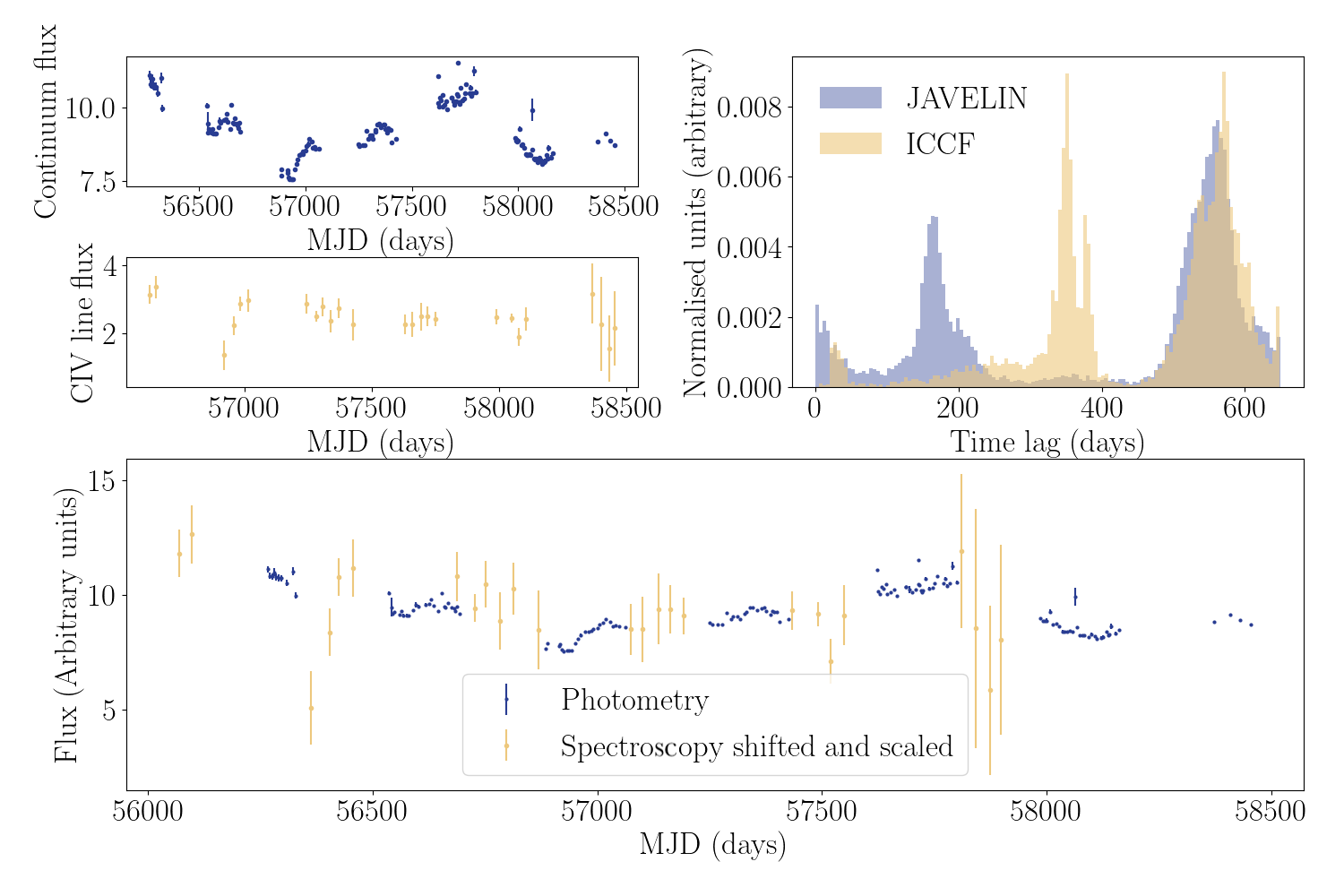}
    \caption[Bronze sample example.]{An example of bronze rated source, DES~J032703.62-274425.27. In the photometric and spectroscopic lightcurves we can see that there is a variation in both lightcurves, however, with a generally lower signal-to-noise than seen in \Cref{fig:goldexample,fig:silverexample}, and very little temporal overlap due to seasonal gaps. This leads to a posterior for the lag with multiple sharp peaks in both \texttt{JAVELIN} and ICCF, however, the most prominent peak in both are the same, giving credibility to this measurement, albeit at lower confidence than the silver and gold samples.  }
    \label{fig:bronzeexample}
\end{figure*}

\section{C\textsc{\textmd{iv}} {\rr Lag} Recovery Statistics}
\label{sec:rec_stats}
In total, there are $305$ \civ sources in the \des/\ozdes sample, although we expect only a fraction of them to provide recoverable lags. Based on projections made at the beginning of the \ozdes campaign by \citet{King2015}, we should expect between $\sim$20\%-30\% successful recoveries from the \civ sample due to a number of limiting factors. Firstly, only a fraction of the AGN are likely to vary significantly during the time of our observations. Secondly, the expected low signal-to-noise ratio in the OzDES data leads to large uncertainties in the individual measurements. Finally, the seasonal gaps in our observing cadence make some time delays difficult to recover {\rr accurately}, particularly those at half yearly intervals, e.g. $180$ days, $540$ days, etc.\ \citep{malik2022}, {\rm where there tend to be more false-positives unless strict selection cuts are made}. 

Using the quality cuts introduced in \citet{penton2021}, we ensure that we only select high-quality lag recoveries.  In Table~\ref{tab:source_stats} we show the number of sources that pass each cut independently and cumulatively.  Detailed descriptions of the cuts can be found in \citet{penton2021}, but in summary:
\begin{enumerate}
    \item {\rr {\bf Consistent with 0 at 3$\sigma$:}} The first cut removes the sources that either have broad and/or noisy posterior distributions that give them large uncertainties consistent with zero lag; this removes 50 sources {\rr (from 305 to 255)}.
    \item {\rr {\bf $|$JAV.$-$ICCF$| < $ 100 days:}} The second cut is based on the disparity between measurements made using ICCF and \texttt{JAVELIN}. This is not a cut that is often made in the literature, however, simulations by \citet{penton2021} showed it to be important; it removes another 145 sources {\rr (from 255 to 110)}. 
    \item The final pair of cuts splits the sample into bronze, silver, and gold.  To make it into each sample both of the following criteria have to be passed to the appropriate level (as listed in \Cref{tab:source_stats}).
    \begin{enumerate}
        \item {\rr {\bf median$-$peak ($<x$ days):}} Do the median and peak of the {\rr JAVELIN lag} distributions agree to within a certain number  {\rr ($x$)} of days?
        \item {\rr {\bf \% in peak ($>y$\%):}} Is there more than a certain percentage {\rr ($y$\%)} of the probability distribution in the {\rr JAVELIN} primary peak?
    \end{enumerate}
\end{enumerate}
These latter criteria penalise sources that have significant likelihood outside the primary (highest) peak.\footnote{We identify the highest peak and integrate the probability distribution between the local minima on either side of the peak.}  We show three examples of recoveries, one each for gold, silver, and bronze, in \Cref{fig:goldexample} to \Cref{fig:bronzeexample}. These cuts result in 6 gold-standard sources, 6 silver, and 17 bronze.  The simulations in \citet{penton2021} show that gold, silver, and bronze all represent high-quality lags --- the mean offset from the true lag ranges from 13 days (gold) to 18 days (bronze) and the false detection percentage ranges from 12\% (gold) to 19\% (bronze).

\begin{table}
\footnotesize
\centering
\begin{tabular}{cc|cc}
        &         & \makecell{\# Pass \\ Each Cut \\ Indep. } & \makecell{\# Pass \\ Each Cut \\ Cumul.} \\ \hline
        & All     &305     & 305          \\
        & Consistent with 0 at $3\sigma$ & 255      & 255          \\
        & $|$JAV.$-$ICCF$|<$100 days      & 131     & 110          \\ \hline
Bronze  & median$-$peak ($<$110 days)  & 146     & 65          \\
        & \% in peak ($>$33\%)    & 85     & 29           \\ \hline
Silver  & median$-$peak ($<$80 days)  & 127     & 57          \\
        & \% in peak ($>$45\%)     & 26     & 12           \\ \hline
Gold    & median$-$peak ($<$65 days)  & 113     & 52          \\
                        & \% in peak ($>60$\%)     & 13     & 6          
\end{tabular}
\caption[Number of sources that remain after each quality cut, both independently and cumulatively.]{Number of sources that remain after each quality cut, both independently and cumulatively. {\rr For example, the $|$JAV$-$ICCF$|$ cut removes 174 sources ($305-131$) if applied independently of the first cut, but only 145 sources ($255-110$) when applied after the first cut.}  Note that the pairs of cuts for gold, silver and bronze indicate the number that pass, firstly, only the cut on the difference between median and peak lag (median$-$peak), and secondly, {\em both} the median$-$peak and the peak proportion cuts for that rating. 
}
\label{tab:source_stats}
\end{table}

\label{sec:potential_increased_recovery}
The cut that has the biggest impact on the final number of recovered sources is the cut on the proportion of the lag posterior contained within the peak of the posterior (only 85 of the 305 AGN pass this criterion at the Bronze level). Given that we find that sources observed at higher redshifts often have lower signal-to-noise due to observational limitations, it is not surprising that a high-redshift sample would contain a significant number of low SNR posterior lag distributions. 

When measuring high quality black hole masses, a strict cut on this parameter is necessary, though it may be possible to instead use this information on a sliding scale. 

\section{Recovering Black Hole Masses}
\label{sec:BHM}
Having collated a set of $29$ time lags, we can now use them to measure the masses of their associated black holes. 
To compute a black hole mass we use the virial equation:
\begin{equation}
        M_{\textrm{BH}} = f\frac{c\,t_{\textrm{lag}} \Delta V^2}{G},
        \label{eq:MBH}
    \end{equation}
where $t_{\textrm {lag}}$ denotes the time lag between the central SMBH and the BLR, 
so that $c t_{\textrm {lag}}$ is an estimate of the BLR radius; $\Delta V$ denotes the BLR velocity dispersion as measured from the reverberating emission line; and $f$ represents the `virial factor,'  an empirical correction factor that accounts for the mean kinematic properties of all AGN including unknown inclination, internal kinematics, and other undetermined properties of the internal regions of the AGN. Multiple estimates for the virial factor are discussed in the literature,  based on different data sets and different methodologies. In this work, we use the virial factor of $f=4.47\pm1.1$ 
based on the work by \citet{2015ApJ...801...38W}, as it is the most widely used -- however, it is important to note that this was derived based on measurements for \hbeta. 
The uncertainty on this measurement is quite large, often being the dominant source of uncertainty in the mass measurements.  

To measure the BLR velocity dispersion for each source we measure the linewidth of the \civ line. This is done by measuring the linewidth of the mean spectrum; however, this does come with limitations. There is some evidence showing that there are non-negligible contributions from outflows in the line profile of some \civ sources. \citet{2012ApJ...759...44D} showed that high velocity outflows from the central region of AGN may cause widening on the blue side of the \civ line. This widening would make linewidth measurements using the mean spectrum possibly inaccurate since these outflows are not part of the orbital motion of the BLR clouds. This could be mitigated by using the RMS of the spectrum, thus showing only the parts of the line profile that are changing with time. 
Unfortunately, the signal-to-noise ratio of the \ozdes spectra are insufficient to reliably measure linewidths of the RMS spectrum. 
However, for single epoch mass measurements, the majority of the outflows that have been observed cause shifts of less than $1\sigma$ in the velocity measurement. Therefore, this effect would be sub-dominant to the uncertainties in the virial factor and the lag uncertainty. 

Four of our 29 sources had successful lag measurements but their \civ\ line suffered from significant absorption features, preventing an accurate measure of line dispersion (see the first, second, fourth, and tenth spectra from the top\footnote{DES~J025105.13-001732.01; DES~J024514.93-004101.83; DES~J022330.15-043004.09; DES~J025100.64+001707.38} in Figure~\ref{fig:all_good_spectra}).  As a result we are left with 25 black hole mass measurements, which are presented in {\rr Appendix} Table~\ref{tab:BHM_results}. 
Our stringent selection criteria mean that this sample is very robust.  We find an intrinsic scatter in our radius-luminosity relation 2.5 times smaller than that in the SDSS sample ($\sim0.2$ dex vs $\sim0.5$ dex); see Figure~7 of \citet{McDougall2025_OzDES}.   

To assess how supermassive black holes have evolved over cosmic time we will require many more sources. To achieve this, 
McDougall et al.~(in prep) will leverage the results presented here to measure black hole masses of a larger sample.  While we have only a limited number of reverberation mapped sources, there are thousands of AGN in the \ozdes catalogue for which we have photometry and spectroscopy, just not over as long a time period.  All of these represent a possible black hole mass measurement using the $R-L$ relation combined with a single-epoch spectroscopic velocity measurement. Future analyses will be able to apply these $R-L$ relations to even larger samples, such as those from SDSS, the Dark Energy Spectroscopic Instrument (DESI), and 4MOST \citep{2019Msngr.175...58S}. 

\begin{figure}
    \centering
    \includegraphics[width=84mm]{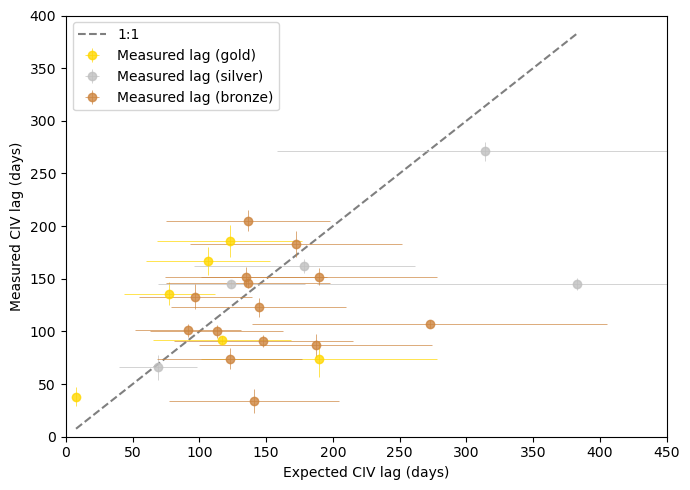}
    \caption[The measured rest frame lags compared to the expected measurements for each source based on historic $R-L$ relation estimates.]{The distribution of recovered rest frame lags and redshifts compared to the expected distribution of the OzDES sample based on the previous $R-L$ relation estimate of \citet{Grier2019}. 
    }
    \label{fig:restframe_v_redshift}
\end{figure}

\begin{figure*}
    \centering
    \includegraphics[width=140mm]{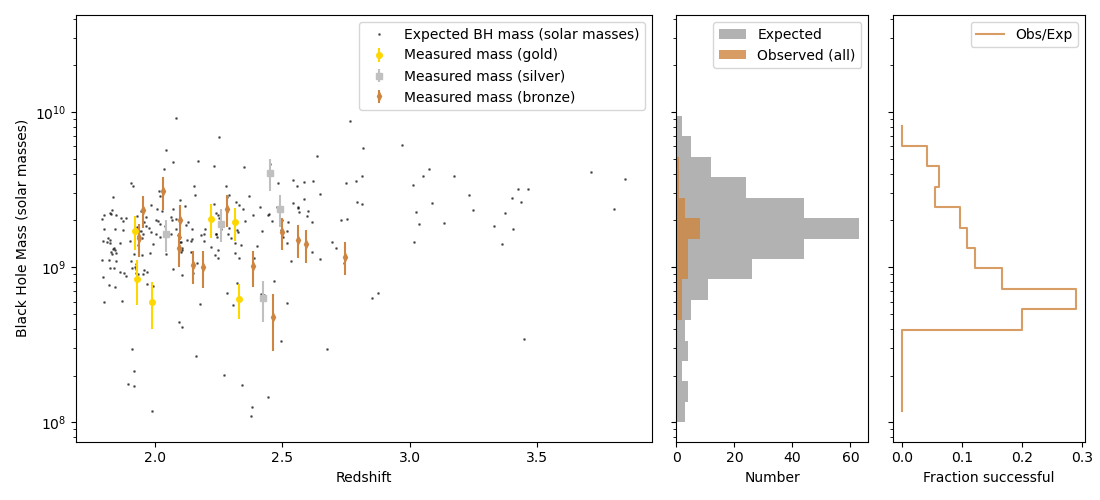}
    \caption[Comparison of predicted distribution of black hole masses for the whole OzDES \civ\ sample compared to the successfully measured sample of the black hole masses.]{A comparison of the distribution of predicted black hole {\rr (BH)} masses {\rr for the whole OzDES sample} with the distribution of successfully measured black hole masses (in units of solar mass).  {\rr The prediction uses the \citet{Grier2019} $R-L$ relation with $f=4.47\pm1.1$ from \citet{2015ApJ...801...38W}.}  {\rr The left panel shows the black hole masses plotted vs redshift, with the measured data coloured by its quality. The central panel shows a histogram of predicted masses compared to the successfully measured (observed) masses.  The right panel shows the fraction of successful lags as a function of mass, showing} the generally smaller masses of the black holes that had successful measurements relative to the input sample.  Note also that we find a low recovery rate above a redshift $z=2.8$, with no recoveries out of a possible $\sim$30, and no gold measurements above $z=2.35$. This is likely due to low intrinsic variability of high-luminosity sources, the low signal-to-noise of high-redshift measurements, and time limit due to the maximum survey duration.}
    \label{fig:BHM_comp}
\end{figure*}

\section{Selection effects, demographics, and the $R-L$ relation}

\label{sec:selection}

In order to infer properties of the population of AGN as a function of redshift, one needs to consider the selection effects that impact the sample. The selection effects occur at two stages.  Firstly, did \ozdes observe a representative sample of AGN? Secondly, of the observed AGN, which had successful{\rr ly measured} lags? 

As a means of contrasting our mass measurements with a priori estimates, we use an $R-L$ relation from the literature \citep{Grier2019} to predict time lags for our entire sample, and combine that with the measured velocity dispersion, to give an estimate of the black hole mass.  This allows us to estimate the distribution of lags and SMBH masses that are expected for the \ozdes sample.\footnote{To derive lags, \citet{Grier2019} use the variations of luminosity $\log\lambda L_{\lambda}(1350\AA)$ in ergs\,s$^{-1}$ compared to the variability of the RMS spectrum of \civ.}
In \Cref{fig:restframe_v_redshift} we compare the 29 measured lags with the predicted lags inferred from the $R-L$ relation \citep{Grier2019}, 
\begin{equation} \log R_{\rm BLR} ({\rm light-days}) = \beta + \alpha \log_{\rm 10}\frac{\lambda L_{\lambda}(1350\AA)}{10^{44}{\rm erg}^{-1}} \pm \epsilon,\label{eq:RL}\end{equation}
where $R_{\rm BLR}$ is the radius of the broad line region in light-days, $\beta=1.15\pm0.008$, $\alpha=0.51\pm0.05$, and the intrinsic scatter is $\epsilon=0.15\pm0.03$. Our results are consistent with the predictions, given the large uncertainties in the predictions.\footnote{Uncertainties ($\sigma$) in the $R-L$ relation (Equation~\ref{eq:RL}) are given by the following, writing $\mathcal{L}=\log_{\rm 10}\frac{\lambda L_{\lambda}(1350\AA)}{10^{44}{\rm erg}^{-1}}=\log_{\rm 10}(\lambda L_{\lambda}(1350\AA){\rm erg})-44$,
\begin{equation} 
\sigma_{R_{\rm BLR}} = R_{\rm BLR}\ln(10)\sqrt{\sigma_\beta^2 + \sigma_\alpha^2\mathcal{L}^2+\sigma_{\mathcal{L}}^2\alpha^2},
\end{equation}
and uncertainties in the black hole mass are propagated as,
\begin{equation}
    \sigma_{M_{\rm BH}}=\frac{f R_{\rm BLR} \Delta V^2}{G}\sqrt{\left(\frac{\sigma_f}{f}\right)^2+\left(\frac{\sigma_{R_{\rm BLR}}}{R_{\rm BLR}}\right)^2+4\left(\frac{\sigma_{\Delta V}}{\Delta V}\right)^2}.
\end{equation}
}

Figure~\ref{fig:BHM_comp} shows the distribution of our black hole mass measurements, compared to the masses one would expect based on the $R-L$ relation from Equation~\ref{eq:RL} and Equation~\ref{eq:MBH} with $f=4.47\pm1.1$ \citep{2015ApJ...801...38W}.

We see the bulk of our successful black hole mass recoveries lie at relatively low redshift and that our success rate is better for lower black hole masses. There are several reasons for this:
\begin{itemize}
    \item {\bf Cadence and Seasonal Gaps:} Short lags are difficult to constrain reliably given the monthly cadence of our observations, and the 6-8 month gaps between seasons.  As discussed in Malik et al. (2022), filling in the seasonal gaps increases the chances of successfully measuring lags.
    \item {\bf Survey duration:} Time dilation lengthens the observer-frame duration of lags, which makes it difficult to observe very long lags at high redshift, as observed lags can become longer than the maximum duration of the survey.  This has an important impact on the $R-L$ relation (see Section~\ref{sec:biases}).
    \item  {\bf Variability time scale}: Sources with longer lags indicate a larger BLR, which is generally correlated with longer-term variability \citep{Lu_2019, Tarrant_2025}, meaning they can be constrained even when there are large gaps in the data. 
    \item {\bf Variability amplitude}: Smaller black holes, contained within lower luminosity AGN, tend to be more variable sources {\rr \citep{MacLeod2012}}. These sources will have lightcurves with higher variability and would therefore be more likely to be recovered.
    \item {\bf Signal-to-noise:} High-redshift targets tend to have lower signal-to-noise than low-redshift targets. 
\end{itemize}

These selection effects can be seen in our black hole mass distribution. 
In {\rr the right panels of} \Cref{fig:BHM_comp} we show histograms of the black hole masses that we have measured through RM  compared to those predicted by the $R-L$ relation.
This comparison shows that, on average, OzDES successfully recovered lags {\rr more efficiently} for the sources with low{\rr er} black hole masses. This is primarily due to the smaller black holes having higher variability, and larger black holes hitting the duration limit of the survey (see Section~\ref{sec:biases}).  

Both of these reasons are selection effects, and are likely not indicative of individual black holes being less massive than expected.  
However, this does indicate that it is difficult to measure the very large black hole masses above $10^{10}M_{\odot}$ with the survey structure and length of \ozdes. A longer baseline and/or deeper spectroscopy would be required to measure these gargantuan black holes. 

One final feature of importance is that many of the \civ observer-frame lags are expected to be in seasonal gaps, because at $2\lesssim z\lesssim 3$ time dilation shifts a rest-frame lag of 50 days to $\sim$150 to 200 days in the observer frame. We found that current lag recovery methods may be overly sensitive to aliasing effects due to seasonal gaps in the light curves.    In future work we aim to use the new LITMUS program to do AGN reverberation mapping \citep{McDougall2025_LITMUS}, as this significantly improves on JAVELIN's method of exploring the parameter space, reducing the impact of aliasing.

\begin{figure}[t]
         \centering
         \includegraphics[width=90mm]{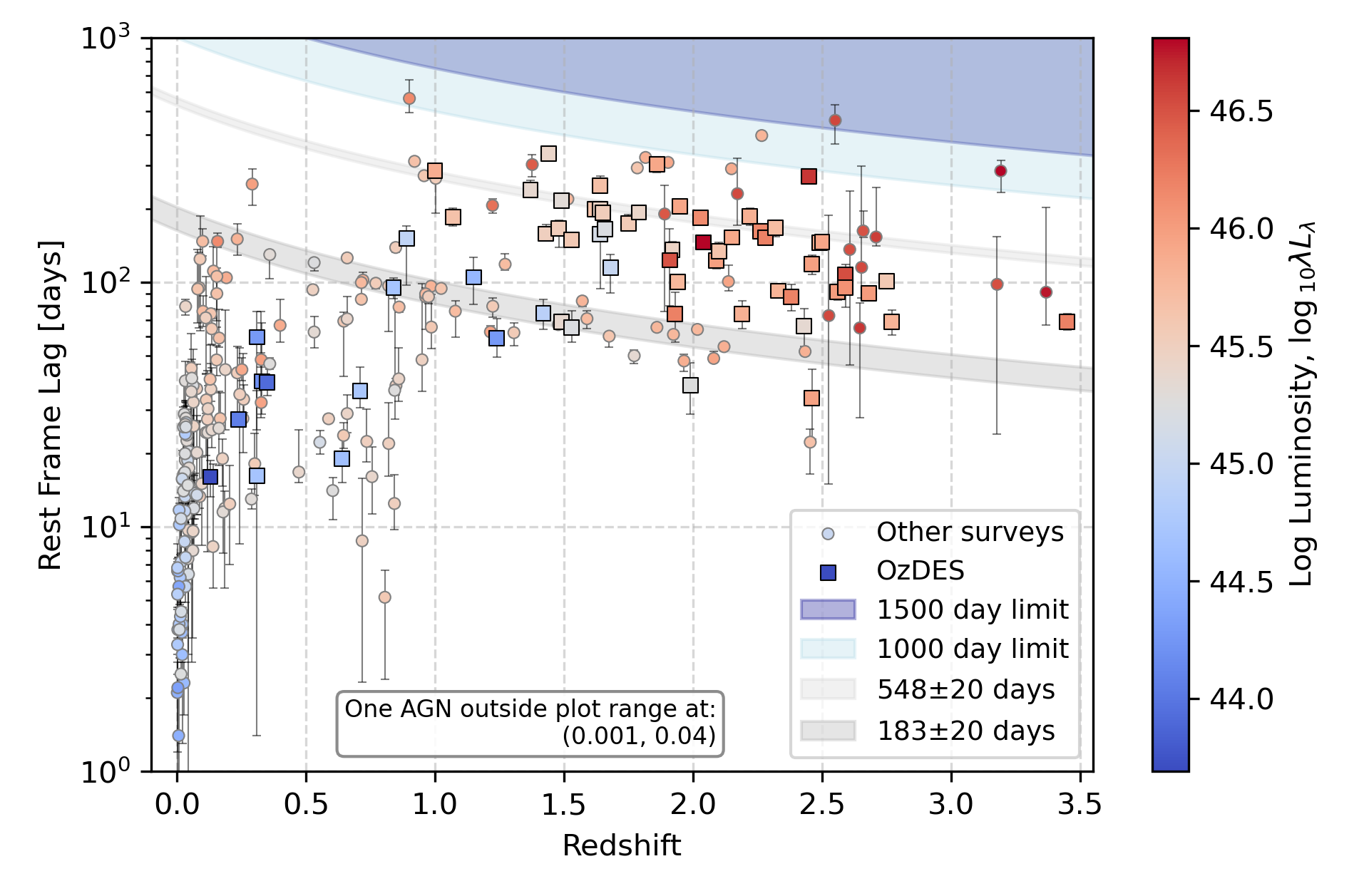}
         \caption[Rest frame lag vs redshift coloured by the luminosity of the AGN.  Shading shows the approximate upper limit of possible lag recoveries given typical survey durations (1000 and 1500 days).]{Compilation of data from the literature and this paper, showing rest-frame lags as a function of redshift, and coloured by the luminosity of the AGN.  Blue shading shows the approximate upper limit of possible lag recoveries given typical survey durations (observer-frame lags of 1000 and 1500 days).  Grey shading shows the approximate position of seasonal gaps. 
         There is a decrease in the highest measurable rest-frame lag as redshift increases. This impacts \civ\ measurements the most, and is an important selection effect for the \civ\ $R-L$ relation (see Figure~\ref{fig:Lag_lum_bias}). Data are from the compilation in \citet{McDougall2025_OzDES}, {\rr which includes \citet{CIV_Peterson_2005, CIV_Rosa_2015, Grier2019, CIV_Kaspi_2021, SDSS-Shen_2024} in addition to the OzDES measurements.}} 
         \label{fig:Lag_red_bias}
\end{figure}

\subsection{Selection Effects' Impact on the Radius-Luminosity Relation}\label{sec:biases}

Measuring black hole masses is only one of the applications of this data set. In addition we can construct a scaling relation based on the measured radius of the broad line region and the luminosity of the AGN.  In \citet{McDougall2025_OzDES} we re-derive the $R-L$ relation for all emission lines used by OzDES, so we defer to that paper for the new \civ\ $R-L$ relation both from OzDES and in combination with existing surveys.\footnote{In an early version of this work, which appears in the thesis of the lead author \citep{PentonPhD2023}, you can find some additional analyses that include a preliminary \civ\ $R-L$ relation, and an analysis including an additional 65 OzDES AGN measurements weighted by the proportion of their posterior contained within the primary likelihood peak.}

Here we look at one important feature of the \civ\ sample that impacts the $R-L$ relation, namely the selection effect that arises due to survey duration.  The observed duration of events is time-dilated by a factor of $1+z$ relative to the duration at emission \citep[recent measurements from quasars and DES supernovae respectively include][]{Lewis2023,White2024}. Therefore, a lag of one year in the rest frame would appear as three years for a $z=2$ object.  Due to the high redshift of the \civ\ sample, it is more affected by time-dilation than \hbeta\ or \mgii.

While exploring the $R-L$ relation in 
\citet{PentonPhD2023}, we discovered a discrepancy between the $R-L$ relation constructed using the literature data points and that using only the data from this work. At first glance it appears that the high-luminosity sample of \ozdes has a shallower slope than the literature data that extend to lower luminosities. 
There are multiple possible reasons for this distinction between low and high luminosity regimes. The first possibility is that there is an intrinsic difference between the sources in the different regimes. 
If there were an evolutionary difference between the populations this could cause the disparity in the relations. On the other hand, there are possible observational biases that would prevent recoveries in certain parts of the $R-L$ relation. For example, if time lags on certain time scales are less likely to be successfully recovered, this could skew the $R-L$ relation.  It is this selection effect we believe to be at play in the OzDES data. 

In \Cref{fig:Lag_red_bias} we plot lag vs redshift, colour coding the points by their luminosity.  For this plot we include high-quality data from the literature as compiled in our companion paper \citep{McDougall2025_OzDES}, including data from \hbeta\ and \mgii\ lines \citep{CIV_Peterson_2005, CIV_Rosa_2015, Grier2019, CIV_Kaspi_2021, SDSS-Shen_2024}.  It is clear that higher luminosities tend to come from sources at higher redshifts.

\begin{figure}[t]
         \centering
         \includegraphics[width=80mm]{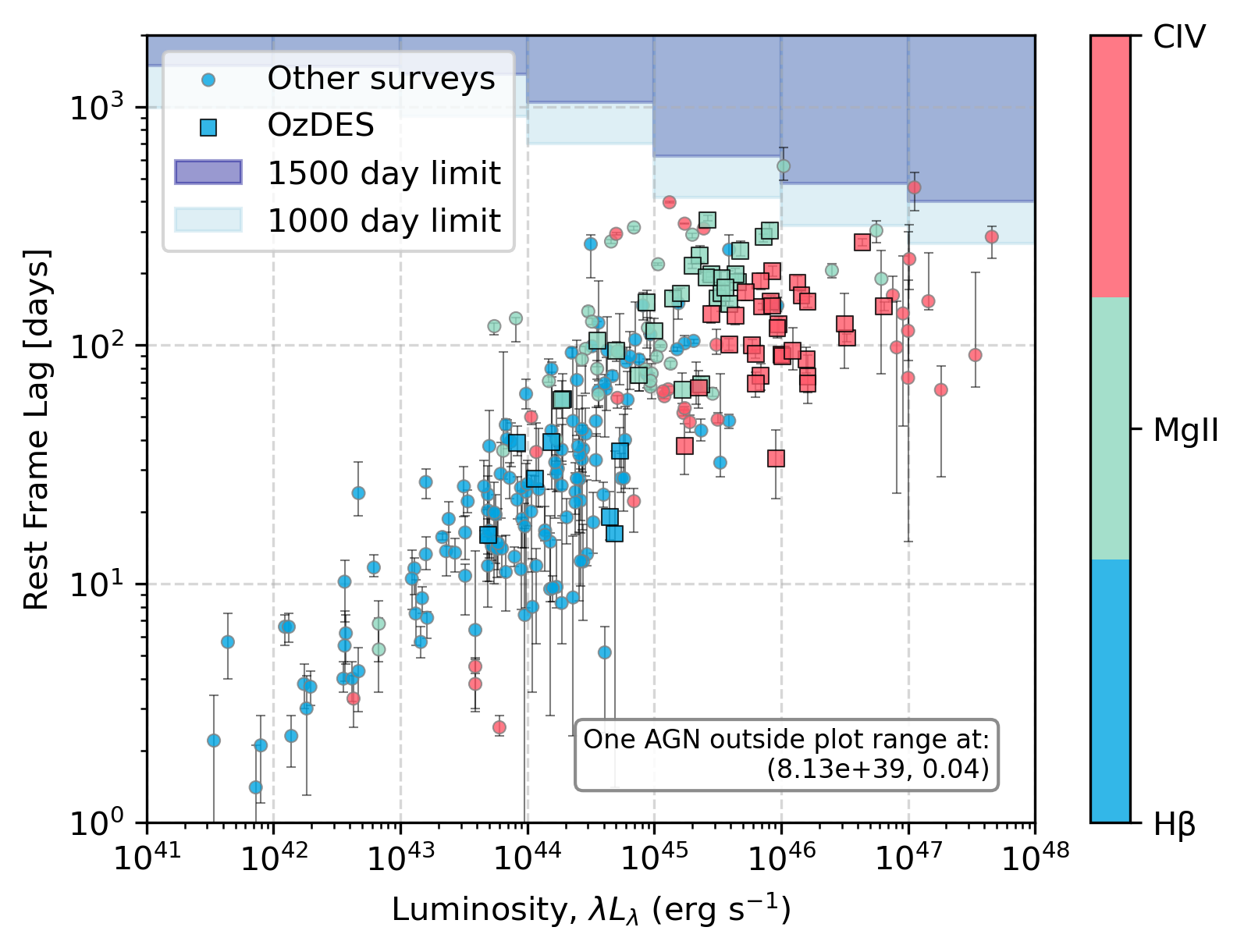}
         \caption[Estimated upper limit of possible recoveries given the average redshift in four luminosity bins.]{Compilation of data from the literature and this paper, showing rest frame lag vs luminosity.  Shaded regions show the estimated upper limit of possible recoveries given the duration of typical large-scale surveys (averaging the limit shown in Figure~\ref{fig:Lag_red_bias} in luminosity bins).  Note that at higher luminosities ($L\gtrsim 10^{46}$~erg\,s$^{-1}$) the areas that are excluded due to survey duration start to impact regions where we expect data. Thus there may be a bias towards points on the lower side of the $R-L$ relation at higher luminosities.}
         \label{fig:Lag_lum_bias}
\end{figure}

A ramification of this can also be seen in \Cref{fig:Lag_red_bias}, where we draw contours of the approximate rest-frame lags that can be measured as a function of redshift.  One cannot recover a lag longer than the survey duration, and ideally one wants the observing window to be significantly longer than the lag.  So these contours (at 1000 and 1500 days in the observer frame) act as approximate upper limits to the lags we could detect (OzDES spanned approximately 2150 days from the first to last observation).  

This is compounded at high redshift by the fact that high-luminosity sources also have intrinsically longer lags (as seen in the $R-L$ relation). 
The cutoffs in \Cref{fig:Lag_red_bias} can therefore be translated into cutoffs that could adversely affect the $R-L$ relation. Since there is not a direct correlation between luminosity and redshift, we have binned the data into {\rr seven} luminosity bins, calculating the average `maximum lag' measurable for the AGN in each of those bins. 
The result of this can be seen in \Cref{fig:Lag_lum_bias}{\rr , which includes the same data sets as Figure~\ref{fig:Lag_red_bias}}.  

There is a clear ceiling where one would expect to see longer duration lags, but they are not detectable because the duration of the survey was insufficient to see the very delayed spectral response to the continuum variation. This is an important selection effect we need to correct for when deriving the $R-L$ relationship for \civ
\citep[see][]{McDougall2025_OzDES}.  

Moving into the future, this bias will need to be carefully parameterised if we are to accurately derive the $R-L$ relation using even larger samples of high-redshift data. One can mitigate this effect by extending RM surveys to longer observational baselines, enabling them to recover longer rest-frame lags successfully.  Such data will be necessary to conclusively rule out a flattening of the slope of the \civ $R-L$ relation.  Notably, the DES supernova fields we used this AGN monitoring campaign continue to be monitored with {\rr the same instrument as DES (DECam)} by the {\rr High-quality Extragalactic Legacy-field Monitoring (HELM)} program \citep{Zhuang2024}, resulting in a photometric sequence that spans more than a decade.   

\section{Conclusions}\label{sec:conclusion}

We recover 29 high-quality \civ lags from observing $305$ AGN over six years with weekly photometric cadence and monthly spectroscopic cadence with the \des/\ozdes program.  
Using these lags in combination with linewidth velocity dispersion measurements, we measure masses of the central supermassive black holes for 25 of these AGN. 

We note that selection effects impact the demographics of the AGN in this sample, which occupies the upper edges of our survey's window function in redshift-luminosity space. Seasonal gaps and the total length of the survey are limiting factors in RM lag recovery, particularly at high redshift where observed lags and timescales of variations are longer. In particular, we identify that building an $R-L$ relationship for the \civ line requires one to take into account the selection effect that the duration of the survey imposes on the maximum recoverable lag. 

Future surveys may allow us to probe different regions of luminosity-redshift space, providing more reliable anchors for the $R-L$ relationship. Industrial-scale surveys such as \ozdes offer a wealth of sources. With the high inherent scatter and the difficulty in constraining the $R-L$ relationship for \civ, there is good motivation to improve the breadth and density of \civ RM sources to properly characterise this behaviour.

\vspace{-5mm}\section*{Contribution Statement}
 
Project conception and coordination: TD, CL, PM; Analysis, programming, calculations: AP, ZY, HM, TD; Writing: AP, HM, TD; Figures: AP, TD; Editing: CL, PM, ZY, RS; Data generation and/or curation: all authors.

\vspace{-5mm}\section*{Acknowledgements}
{\footnotesize
AP, UM, and HM are supported by the Australian Government Research Training Program (RTP) Scholarship. TMD acknowledges support for early stages of this project from an Australian Research Council (ARC) Laureate Fellowship (project number FL180100168) and for later stages from the ARC Centre of Excellence for Gravitational Wave Discovery, OzGrav (CE230100016).  PM and ZY were supported in part by the United States National Science Foundation under Grant No. 161553 to PM. PM also acknowledges support from the United States Department of Energy, Office of High Energy Physics under Award Number DE-SC-0011726. 

Parts of this research were carried out on the traditional lands of the Ngunnawal and Ngambri peoples. This work makes use of data acquired at the Anglo-Australian Telescope, under program A/2013B/012. We acknowledge the Gamilaraay people as the traditional owners of the land on which the AAT stands. We pay our respects to their elders past, present, and emerging.

This analysis used \textsc{NumPy} \citep{harris2020array}, \textsc{Astropy} \citep{astropy:2013, astropy:2018}, and \textsc{SciPy} \citep{2020SciPy-NMeth}. Plots were made using \textsc{Matplotlib} \citep{Hunter:2007}. This work has made use of the SAO/NASA Astrophysics Data System Bibliographic Services.

Funding for the DES Projects has been provided by the U.S. Department of Energy, the U.S. National Science Foundation, the Ministry of Science and Education of Spain, 
the Science and Technology Facilities Council of the United Kingdom, the Higher Education Funding Council for England, the National Center for Supercomputing 
Applications at the University of Illinois at Urbana-Champaign, the Kavli Institute of Cosmological Physics at the University of Chicago, 
the Center for Cosmology and Astro-Particle Physics at the Ohio State University,
the Mitchell Institute for Fundamental Physics and Astronomy at Texas A\&M University, Financiadora de Estudos e Projetos, 
Funda{\c c}{\~a}o Carlos Chagas Filho de Amparo {\`a} Pesquisa do Estado do Rio de Janeiro, Conselho Nacional de Desenvolvimento Cient{\'i}fico e Tecnol{\'o}gico and 
the Minist{\'e}rio da Ci{\^e}ncia, Tecnologia e Inova{\c c}{\~a}o, the Deutsche Forschungsgemeinschaft and the Collaborating Institutions in the Dark Energy Survey. 

The Collaborating Institutions are Argonne National Laboratory, the University of California at Santa Cruz, the University of Cambridge, Centro de Investigaciones Energ{\'e}ticas, 
Medioambientales y Tecnol{\'o}gicas-Madrid, the University of Chicago, University College London, the DES-Brazil Consortium, the University of Edinburgh, 
the Eidgen{\"o}ssische Technische Hochschule (ETH) Z{\"u}rich, 
Fermi National Accelerator Laboratory, the University of Illinois at Urbana-Champaign, the Institut de Ci{\`e}ncies de l'Espai (IEEC/CSIC), 
the Institut de F{\'i}sica d'Altes Energies, Lawrence Berkeley National Laboratory, the Ludwig-Maximilians Universit{\"a}t M{\"u}nchen and the associated Excellence Cluster Universe, 
the University of Michigan, NSF NOIRLab, the University of Nottingham, The Ohio State University, the University of Pennsylvania, the University of Portsmouth, 
SLAC National Accelerator Laboratory, Stanford University, the University of Sussex, Texas A\&M University, and the OzDES Membership Consortium.

Based in part on observations at NSF Cerro Tololo Inter-American Observatory at NSF NOIRLab (NOIRLab Prop. ID 2012B-0001; PI: J. Frieman), which is managed by the Association of Universities for Research in Astronomy (AURA) under a cooperative agreement with the National Science Foundation.

The DES data management system is supported by the National Science Foundation under Grant Numbers AST-1138766 and AST-1536171.
Data access is enabled by Jetstream2 and OSN at Indiana University through allocation PHY240006: Dark Energy Survey from the Advanced Cyberinfrastructure Coordination Ecosystem: Services and Support (ACCESS) program, which is supported by U.S. National Science Foundation grants 2138259, 2138286, 2138307, 2137603, and 2138296.
The DES participants from Spanish institutions are partially supported by MICINN under grants PID2021-123012, PID2021-128989 PID2022-141079, SEV-2016-0588, CEX2020-001058-M and CEX2020-001007-S, some of which include ERDF funds from the European Union. IFAE is partially funded by the CERCA program of the Generalitat de Catalunya.

We  acknowledge support from the Brazilian Instituto Nacional de Ci\^encia
e Tecnologia (INCT) do e-Universo (CNPq grant 465376/2014-2).

This document was prepared by the DES Collaboration using the resources of the Fermi National Accelerator Laboratory (Fermilab), a U.S. Department of Energy, Office of Science, Office of High Energy Physics HEP User Facility. Fermilab is managed by Fermi Forward Discovery Group, LLC, acting under Contract No. 89243024CSC000002.
}

\def \aap{A\&A}
\def \aaps{A\&AS}
\def \ag{Astron. Geophys.}
\def \aj{AJ}
\def \ajp{Am. J. Phys.}
\def \al{Astron. Lett.}
\def \ap{Appl. Phys.}
\def \apj{ApJ}
\def \apjl{ApJ}
\def \apjl{ApJ Lett.}
\def \apjs{ApJS}
\def \apss{Astrophys. and Space Science}
\def \araa{ARA\&A}
\def \arns{Annu. Rev. Nuc. Sci.}
\def \asp{Astron. Soc. Pac.}
\def \azh{Astronomicheskii Zhurnal}
\def \baas{BAAS}
\def \baps{Bull. Am. Phys. Soc.}
\def \bist{Bull. Inf. Sci. Tech.}
\def \ca{Comments on Astrophys.}
\def \cqg{Class. Quantum Gravity}
\def \epjc{Euro. Phys. J. C}
\def \grg{Gen. Relativ. Gravitation}
\def \ijmpd{Int. J. Mod. Phys. D}
\def \jhep{J. High Energy Phys.}
\def \jms{J. Molecular Spectrosc.}
\def \jos{J. Opt. Soc. Am.}
\def \jcap{J. Cosmo. Astropart. Phys.}
\def \josb{J. Opt. Soc. Am. B}
\def \jpcrd{J. Phys. Chem. Ref. Data}
\def \jpcrds{J. Phys. Chem. Ref. Data Suppl.}
\def \jqsrt{J. Quant. Spectrosc. Radiat. Transfer}
\def \jtp{J. Technical Phys.}
\def \met{Metrologia}
\def \mnras{MNRAS}
\def \mpla{Mod. Phys. Lett. A}
\def \nat{Nature}
\def \nature{Nature}
\def \npb{Nucl. Phys. B}
\def \nsrds{Natl. Stand. Rel. Data Ser.}
\def \nw{Naturwiss.}
\def \pasa{Publ. Astron. Soc. Austral.}
\def \pasp{PASP}
\def \pawk{Preuss. Akad. Wiss. K}
\def \plb{Phys. Lett. B}
\def \phd{PhD thesis}
\def \physrep{Physics Reports}
\def \pr{Phys. Rev.}
\def \pra{Phys. Rev. A}
\def \prb{Phys. Rev. B}
\def \prc{Phys. Rev. C}
\def \prd{Phys. Rev. D}
\def \prep{in preparation}
\def \prl{Phys. Rev. Lett.}
\def \prsa{Proc. R. Soc. A}
\def \psc{Phys. Scr.}
\def \ptp{Progress Theor. Phys.}
\def \ptrsla{Phil. Trans. R. Soc. London}
\def \qjras{Quart. J. R. Astron. Soc.}
\def \rmp{Rev. Mod. Phys.}
\def \rpp{Rep. Prog. Phys.}
\def \sovast{Soviet Astronomy}
\def \spu{Sov. Phys. Uspekhi}
\def \ssr{Space Sci. Rev.}
\def \tms{Trans. Math. Software}

\def \cup{Cambridge U. Press}
\def \cupadr{Cambridge U.K.}
\def \pup{Princeton U. Press}
\def \pupadr{Princeton, U.S.A.}

\def \etal{et al.}

\vspace{-5mm}\bibliographystyle{mnras}
\bibliography{thesis, bib_CIVsources}

\vspace{-3mm}
\subsection*{Affiliations}
{\footnotesize
$^{1}$School of Mathematics and Physics, University of Queensland,  Brisbane, QLD 4072, Australia, $^{2}$Kavli Institute for Particle Astrophysics \& Cosmology, P. O. Box 2450, Stanford University, Stanford, CA 94305, USA, $^{3}$The Research School of Astronomy and Astrophysics, Australian National University, ACT 2601, Australia, $^{4}$Center for Cosmology and Astro-Particle Physics, The Ohio State University, Columbus, OH 43210, USA, $^{5}$Department of Astronomy, The Ohio State University, Columbus, OH 43210, USA, $^{6}$Centre for Gravitational Astrophysics, College of Science, The Australian National University, ACT 2601, Australia, $^{7}$Sydney Institute for Astronomy, School of Physics, A28, The University of Sydney, NSW 2006, Australia, $^{8}$INAF-Osservatorio Astronomico di Trieste, via G. B. Tiepolo 11, I-34143 Trieste, Italy, $^{9}$Laborat\'orio Interinstitucional de e-Astronomia - LIneA, Av. Pastor Martin Luther King Jr, 126 Del Castilho, Nova Am\'erica Offices, Torre 3000/sala 817, Brazil, $^{10}$Fermi National Accelerator Laboratory, P. O. Box 500, Batavia, IL 60510, USA, $^{11}$Physik-Institut, University of Z\'{u}rich, Winterthurerstrasse 190, CH-8057 Z{\'u}rich, Switzerland, $^{12}$Departamento de F\'{\i}sica Te\'orica, Centro de Astropart\'iculas y F\'isica de Altas Energ\'ias (CAPA), Universidad de Zaragoza, 50009 Zaragoza, Spain, $^{13}$Institute of Cosmology and Gravitation, University of Portsmouth, Portsmouth, PO1 3FX, UK, $^{14}$University Observatory, LMU Faculty of Physics, Scheinerstr. 1, 81679 Munich, Germany, $^{15}$Department of Physics \& Astronomy, University College London, Gower Street, London, WC1E 6BT, UK, $^{16}$Instituto de Astrofisica de Canarias, E-38205 La Laguna, Tenerife, Spain, $^{17}$Universidad de La Laguna, Dpto. Astrof{\'i}sica, E-38206 La Laguna, Tenerife, Spain, $^{18}$Korea Astronomy and Space Science Institute, 776 Daedeok-daero, Yuseong-gu, Daejeon 34055, South Korea, $^{19}$Institut de F\'{\i}sica d'Altes Energies (IFAE), The Barcelona Institute of Science and Technology, Campus UAB, 08193 Bellaterra (Barcelona) Spain, $^{20}$Kapteyn Astronomical Institute, University of Groningen, Landleven 12 (Kapteynborg, 5419), 9747 AD Groningen, The Netherlands, $^{21}$Hamburger Sternwarte, Universit\"{a}t Hamburg, Gojenbergsweg 112, 21029 Hamburg, Germany, $^{22}$Centro de Investigaciones Energ\'eticas, Medioambientales y Tecnol\'ogicas (CIEMAT), Madrid, Spain, $^{23}$Department of Physics, IIT Hyderabad, Kandi, Telangana 502285, India, $^{24}$California Institute of Technology, 1200 East California Blvd, MC 249-17, Pasadena, CA 91125, USA, $^{25}$Instituto de F\'{i}sica Te\'{o}rica UAM/CSIC, Universidad Aut\'{o}noma de Madrid, 28049 Madrid, Spain, $^{26}$Centre for Astrophysics \& Supercomputing, Swinburne University of Technology, Victoria 3122, Australia, $^{27}$Santa Cruz Institute for Particle Physics, Santa Cruz, CA 95064, USA, $^{28}$Department of Physics, The Ohio State University, Columbus, OH 43210, USA, $^{29}$Lowell Observatory, 1400 Mars Hill Rd, Flagstaff, AZ 86001, USA, $^{30}$Jet Propulsion Laboratory, California Institute of Technology, 4800 Oak Grove Dr., Pasadena, CA 91109, USA, $^{31}$Department of Physics and Astronomy, University of Pennsylvania, Philadelphia, PA 19104, USA, $^{32}$George P. and Cynthia Woods Mitchell Institute for Fundamental Physics and Astronomy, and Department of Physics and Astronomy, Texas A\&M University, College Station, TX 77843,  USA, $^{33}$Universit\'e Grenoble Alpes, CNRS, LPSC-IN2P3, 38000 Grenoble, France, $^{34}$Instituci\'o Catalana de Recerca i Estudis Avan\c{c}ats, E-08010 Barcelona, Spain, $^{35}$Department of Astrophysical Sciences, Princeton University, Peyton Hall, Princeton, NJ 08544, USA, $^{36}$School of Mathematics and Physics, University of Surrey, Guildford, Surrey, GU2 7XH, UK, $^{37}$Observat\'orio Nacional, Rua Gal. Jos\'e Cristino 77, Rio de Janeiro, RJ - 20921-400, Brazil, $^{38}$SLAC National Accelerator Laboratory, Menlo Park, CA 94025, USA, $^{39}$Ruhr University Bochum, Faculty of Physics and Astronomy, Astronomical Institute, German Centre for Cosmological Lensing, 44780 Bochum, Germany, $^{40}$Laboratoire de physique des 2 infinis Ir\`ene Joliot-Curie, CNRS Universit\'e Paris-Saclay, Bât. 100, F-91405 Orsay Cedex, France, $^{41}$Department of Physics and Astronomy, Pevensey Building, University of Sussex, Brighton, BN1 9QH, UK, $^{42}$Physics Department, Lancaster University, Lancaster, LA1 4YB, UK, $^{43}$Computer Science and Mathematics Division, Oak Ridge National Laboratory, Oak Ridge, TN 37831, $^{44}$Center for Astrophysical Surveys, National Center for Supercomputing Applications, 1205 West Clark St., Urbana, IL 61801, USA, $^{45}$Central University of Kerala, Kasaragod, Kerala, India, $^{46}$Berkeley Center for Cosmological Physics, Department of Physics, University of California, Berkeley, CA 94720, US, $^{47}$Lawrence Berkeley National Laboratory, 1 Cyclotron Road, Berkeley, CA 94720, USA
}
\appendix

\section{Data}
The data in Table A1 are available for download from Zenodo: \url{https://doi.org/10.5281/zenodo.20124709}. Spectra are available from AAO Data Central: \url{https://docs.datacentral.org.au/ozdes/}. Light curves and additional information on the complete OzDES reverberation mapping sample are available on Zenodo: \url{https://doi.org/10.5281/zenodo.20120159}. Images and additional metadata are available from the DES data release: \url{https://des.ncsa.illinois.edu/releases/dr2/dr2-access}.

\begin{landscape}
\begin{table}
\centering
{\small 
\begin{tabular}{cccccccccccccc}
\hline
OzDES ID & J2000 & $z$ & \makecell{log$L$(1350\AA) \\ ergs$^{-1}$} & \makecell{Linewidth\\(km\,s$^{-1}$)} & \makecell{Obs Lag \\ (days)} & \makecell{Rest Lag \\ (days)} & \makecell{BH Mass \\ ($10^9$ $M_{\odot}$)} & \makecell{Gap \\  (days)} & \makecell{Peak-Med \\Rating} & \makecell{Peak\\Prop.}& \makecell{Final\\Rating}& \makecell{Exp.\ Lag \\ (days)}  & \makecell{Exp. BHM \\($10^9$ $M_{\odot}$)} \\ \hline 
2938498296 & 022327.85-040119.16 & $1.924$ & $45.45 \pm 00.08$ & $3808 \pm 171$ & $396 \pm 33$ & $136 \pm 11$ & $1.72 \pm 0.43$ & $50$ & gold & $0.68$ & gold & $78 \pm 34$ & $0.98 \pm 0.5$ \\
2938970755 & 022514.39-044700.14 & $1.927$ & $46.21 \pm 00.01$ & $3610 \pm 31$ & $217 \pm 50$ & $74 \pm 17$ & $0.84 \pm 0.27$ & $37$ & gold & $0.945$ & gold & $189 \pm 88$ & $2.1 \pm 1.1$ \\
2938937393 & 022537.03-050109.34 & $1.936$ & $45.77 \pm 00.02$ & $4190 \pm 34$ & $294 \pm 17$ & $100 \pm 6$ & $1.53 \pm 0.36$ & $13$ & gold & $0.346$ & bronze & $113 \pm 50$ & $1.72 \pm 0.87$ \\
2939629447 & 033401.79-265054.28 & $1.949$ & $45.93 \pm 00.02$ & $3605 \pm 77$ & $604 \pm 29$ & $205 \pm 10$ & $2.33 \pm 0.55$ & $61$ & bronze & $0.389$ & bronze & $136 \pm 61$ & $1.54 \pm 0.79$ \\
2925783603 & 025159.70-005159.89 & $1.989$ & $43.47 \pm 00.01$ & $4239 \pm 135$ & $113 \pm 27$ & $38 \pm 9$ & $0.6 \pm 0.2$ & $45$ & gold & $0.822$ & gold & $7.6 \pm 3.4$ & $0.118 \pm 0.061$ \\
2939433380 & 032703.62-274425.27 & $2.019$ & $46.13 \pm 00.01$ & $4398 \pm 43$ & $556 \pm 35$ & $183 \pm 12$ & $3.09 \pm 0.74$ & $19$ & gold & $0.39$ & bronze & $172 \pm 80$ & $2.9 \pm 1.5$ \\
2970913529 & 002959.21-434835.24 & $2.036$ & $46.81 \pm 00.01$ & $3595 \pm 11$ & $442 \pm 18$ & $145 \pm 6$ & $1.64 \pm 0.38$ & $2$ & gold & $0.482$ & silver & $380 \pm 190$ & $4.3 \pm 2.4$ \\
2937538328 & 021921.81-043642.21 & $2.096$ & $45.98 \pm 00.02$ & $3519 \pm 125$ & $379 \pm 29$ & $123 \pm 9$ & $1.33 \pm 0.33$ & $8$ & gold & $0.375$ & bronze & $144 \pm 66$ & $1.55 \pm 0.81$ \\
2938664659 & 021941.16-044100.36 & $2.096$ & $45.64 \pm 00.04$ & $4146 \pm 83$ & $413 \pm 36$ & $133 \pm 12$ & $2 \pm 0.5$ & $19$ & gold & $0.357$ & bronze & $97 \pm 42$ & $1.45 \pm 0.73$ \\
2940923721 & 032829.96-274212.23 & $2.149$ & $45.92 \pm 00.01$ & $2791 \pm 59$ & $480 \pm 27$ & $152 \pm 9$ & $1.03 \pm 0.25$ & $34$ & bronze & $0.355$ & bronze & $135 \pm 61$ & $0.91 \pm 0.47$ \\
2940479063 & 032939.97-284952.40 & $2.193$ & $45.84 \pm 00.01$ & $3929 \pm 138$ & $237 \pm 31$ & $74 \pm 10$ & $1 \pm 0.27$ & $11$ & gold & $0.357$ & bronze & $123 \pm 55$ & $1.64 \pm 0.84$ \\
2938353110 & 022001.63-052216.92 & $2.218$ & $45.84 \pm 00.02$ & $3548 \pm 64$ & $599 \pm 49$ & $186 \pm 15$ & $2.04 \pm 0.5$ & $6$ & gold & $0.63$ & gold & $123 \pm 55$ & $1.34 \pm 0.68$ \\
2970717582 & 003743.89-434715.68 & $2.252$ & $46.16 \pm 00.01$ & $3672 \pm 60$ & $527 \pm 23$ & $162 \pm 7$ & $1.91 \pm 0.45$ & $52$ & silver & $0.529$ & silver & $178 \pm 83$ & $2.1 \pm 1.1$ \\
2940271483 & 033655.83-290218.23 & $2.289$ & $46.21 \pm 00.01$ & $4228 \pm 77$ & $499 \pm 25$ & $152 \pm 8$ & $2.37 \pm 0.56$ & $76$ & gold & $0.364$ & bronze & $189 \pm 88$ & $2.9 \pm 1.6$ \\
2938924759 & 022410.96-050653.95 & $2.314$ & $45.72 \pm 00.02$ & $3647 \pm 57$ & $552 \pm 42$ & $167 \pm 13$ & $1.94 \pm 0.47$ & $3$ & gold & $0.735$ & gold & $106 \pm 47$ & $1.23 \pm 0.62$ \\
2939280338 & 033843.76-294922.54 & $2.323$ & $45.80 \pm 00.02$ & $2781 \pm 198$ & $307 \pm 12$ & $92 \pm 4$ & $0.62 \pm 0.16$ & $3$ & gold & $0.667$ & gold & $117 \pm 52$ & $0.79 \pm 0.41$ \\
2970571450 & 004056.56-431446.40 & $2.384$ & $46.20 \pm 00.01$ & $3655 \pm 44$ & $295 \pm 37$ & $87 \pm 11$ & $1.01 \pm 0.27$ & $5$ & gold & $0.331$ & bronze & $187 \pm 87$ & $2.2 \pm 1.1$ \\
2925793267 & 025102.06-004142.78 & $2.428$ & $45.35 \pm 00.03$ & $3317 \pm 172$ & $227 \pm 40$ & $66 \pm 12$ & $0.63 \pm 0.19$ & $1$ & gold & $0.578$ & silver & $69 \pm 29$ & $0.66 \pm 0.33$ \\
2938968217 & 022354.81-044814.94 & $2.449$ & $46.64 \pm 00.01$ & $4134 \pm 23$ & $934 \pm 31$ & $271 \pm 9$ & $4.04 \pm 0.94$ & $83$ & gold & $0.51$ & silver & $310 \pm 160$ & $4.7 \pm 2.6$ \\
2925733402 & 024511.94-011317.50 & $2.460$ & $45.96 \pm 00.01$ & $4021 \pm 90$ & $116 \pm 37$ & $34 \pm 11$ & $0.48 \pm 0.19$ & $6$ & bronze & $0.336$ & bronze & $141 \pm 64$ & $2 \pm 1$ \\
2925473757 & 025100.64+001707.38 & $2.460$ & $45.97 \pm 00.02$ & - & $409 \pm 37$ & $118 \pm 11$ & - & $71$ & bronze & $0.359$ & bronze & $143 \pm 65$ & - \\
2940830092 & 032640.93-283206.80 & $2.497$ & $45.85 \pm 00.01$ & $4328 \pm 58$ & $507 \pm 15$ & $145 \pm 4$ & $2.37 \pm 0.55$ & $41$ & gold & $0.484$ & silver & $124 \pm 55$ & $2 \pm 1$ \\
2970730486 & 003957.42-434107.92 & $2.502$ & $45.93 \pm 00.01$ & $3636 \pm 57$ & $512 \pm 17$ & $146 \pm 5$ & $1.68 \pm 0.39$ & $19$ & gold & $0.384$ & bronze & $136 \pm 61$ & $1.56 \pm 0.8$ \\
2940361224 & 022259.87-063326.65 & $2.563$ & $46.00 \pm 00.01$ & $4352 \pm 95$ & $325 \pm 23$ & $91 \pm 6$ & $1.5 \pm 0.36$ & $18$ & gold & $0.435$ & bronze & $148 \pm 67$ & $2.4 \pm 1.3$ \\
2971212466 & 003352.72-425452.55 & $2.586$ & $46.52 \pm 00.01$ & $3876 \pm 19$ & $386 \pm 14$ & $107 \pm 4$ & $1.4 \pm 0.33$ & $65$ & bronze & $0.409$ & bronze & $270 \pm 130$ & $3.6 \pm 1.9$ \\
2938405366 & 022330.15-043004.09 & $2.674$ & $45.58 \pm 00.04$ & - & $331 \pm 20$ & $90 \pm 5$ & - & $53$ & gold & $0.349$ & bronze & $90 \pm 39$ & - \\
2938718841 & 022620.86-045946.48 & $2.743$ & $45.59 \pm 00.02$ & $3650 \pm 104$ & $378 \pm 21$ & $101 \pm 6$ & $1.17 \pm 0.28$ & $6$ & gold & $0.425$ & bronze & $91 \pm 40$ & $1.06 \pm 0.53$ \\
2925599608 & 024514.93-004101.83 & $2.777$ & $45.80 \pm 00.02$ & - & $260 \pm 31$ & $69 \pm 8$ & - & $25$ & gold & $0.351$ & bronze & $117 \pm 52$ & - \\
2925437937 & 025105.13-001732.01 & $3.450$ & $46.21 \pm 00.01$ & - & $307 \pm 24$ & $69 \pm 5$ & - & $2$ & gold & $0.537$ & silver & $189 \pm 88$ & - \\
\hline
\end{tabular}
}
\caption[Successful black hole mass measurements]{Data for all 29 successful lag measurements and 25 successfully recovered black hole mass measurements.  Four sources had successful lag measurements yet lack BH masses.  This is because their \civ line suffered from major absorption features, preventing an accurate measure of the linewidth, which we use as a proxy for the velocity dispersion used in Equation~\ref{eq:MBH}. The last two columns show the expected lag and black hole mass measurements, given by Equation~\ref{eq:RL} and Equation~\ref{eq:MBH} respectively.  The electronic version of this table also includes the AGN that did not have successful lags (the background points in Figure 6).}
\label{tab:BHM_results}
\end{table}

\end{landscape}
\end{document}